\documentclass[useAMS]{mn2e} 

\usepackage{amssymb,amsmath,psfig,times,graphicx,longtable}
\def\gsim{ \lower .75ex \hbox{$\sim$} \llap{\raise .27ex \hbox{$>$}} }
\def\lsim{ \lower .75ex\hbox{$\sim$} \llap{\raise .27ex \hbox{$<$}} }

\def\sc{Schwarzschild}
\def\beq{\begin{equation}}
\def\eeq{\end{equation}}

\def\fe{{\it Fermi}}

\def\fg{$F_{\gamma}$}

\def\fr{$F_{\rm r}$}
\def\obs{$F_{\gamma}$--$F_{\rm r}$}
\def\true{$\hat{F_{\gamma}}$--$\hat{F_{\rm r}}$}
\def\lglr{$L_{\gamma}$--$L_{\rm r}$}

\title[$\gamma$--ray/radio flux correlation in blazars]
{The radio/$\gamma$--ray connection in \fe--blazars}  
\author[G. Ghirlanda, G. Ghisellini, F. Tavecchio, L. Foschini, G. Bonnoli]
{G. Ghirlanda\thanks{Email: giancarlo.ghirlanda@brera.inaf.it}, 
G. Ghisellini, F. Tavecchio, L. Foschini, G. Bonnoli \\
INAF -- Osservatorio Astronomico di Brera, Via Bianchi 46, I--23807 Merate, Italy\\
}

\date{Accepted 2010 December 8. In original form 2010 July 15}

\begin{document}  

\maketitle

\begin{abstract}
We study the correlation between the $\gamma$--ray flux  
(\fg), averaged over the first 11 months of \fe\ survey and integrated above 100 MeV, and the 
radio flux density (\fr\ at 20 GHz) 
of \fe\ sources associated with a radio counterpart in 
the AT20G survey. 
Considering the blazars detected in both bands,
the correlation is highly significant 
and has the form \fg$\propto$\fr$^{0.85\pm0.04}$, 
similar for BL Lac and FSRQ sources. 
However, only a small fraction ($\sim$1/15) of the AT20G radio sources 
with flat radio spectrum, are detected by \fe.
To understand if this correlation is real, we examine the selection effects
introduced by the flux limits of both the radio and $\gamma$--ray surveys, 
and the importance of variability of the $\gamma$--ray flux.
After accounting for these effects, we find that the radio/$\gamma$--ray flux
correlation is real, but its slope is
steeper than the observed one, i.e. \fg$\propto$\fr$^\delta$ with $\delta$ 
in the range $1.25-1.5$.
The observed \obs\ correlation and the fraction of radio sources detected by
\fe\ is reproduced assuming a long term
$\gamma$--ray flux variability  following a log--normal probability distribution with standard 
deviation $\sigma\ge$0.5 
(corresponding to $F_\gamma$ varying by at least a factor 3).
Such a variability  is compatible, even if not necessarily equal, with what observed when comparing, 
for the sources in common, the EGRET and the \fe\ $\gamma$--ray fluxes (even if the \fe\ fluxes are averaged over $\sim$1 year).
Another indication of variability is the non detection of 12 out of 66 EGRET blazars by \fe, despite its higher sensitivity.
We also study the strong linear correlation between the $\gamma$--ray and the radio luminosity 
of the 144 AT20G--\fe\ associations with known redshift and show, through partial correlation analysis, 
that it is statistically robust. 
Two  possible implications of these correlations are discussed: the contribution 
of blazars to the extragalactic $\gamma$--ray background and the prediction of blazars that might 
undergo extremely high states of $\gamma$--ray emission in the next years. 
\end{abstract}

\section{Introduction}

The Large Area Telescope (LAT) on board the \fe\ satellite 
(Atwood et al. 2009) detected 1451 sources (1FGL catalogue) in 
the $\gamma$--ray band above 100 MeV with a significance $\ge$4.5$\sigma$ 
during its first 11 months survey (Abdo et al. 2010, A10 hereafter): 
831 out of 1451 are classified as AGN 
(Abdo et al. 2010a)\footnote{http://heasarc.gsfc.nasa.gov/W3Browse/fermi/fermilac.html}.

We cross correlated (Ghirlanda et al. 2010, G10 hereafter)
the \fe\ 1FGL catalogue with a complete flux limited sample of radio 
sources detected by the Australia Telescope Compact Array (ATCA) in 
a survey conducted at 20 GHz with a flux density limit $\ge$40 mJy (Murphy et al., 2010). 
The cross correlation led to identify highly probable (association 
probability $>$80\%) radio counterparts for 230 1FGL sources, i.e. the 1FGL--AT20G associations 
hereafter (see also Mahony et al. 2010). 
222 of these are already classified in the first LAT 
AGN Catalogue (1LAC -- Abdo et al. 2010a)  as BL Lacs (54), FSRQs (112), 
candidate blazars of unknown class (46) or other type of AGNs (10). 
Among the generic class of "AGN" there are different source types: starburst galaxies  
(NGC 253, M82 -- Abdo et al., 2010b), 
starburst/Seyfert 2 (NGC 4945 -- Lenc  \& Tingay, 2009) 
a Low--Excitation FRI radio galaxy (PKS 0625--35 -- Gliozzi  et al., 2008) 
a High--Excitation FRI radio galaxy (Cen A -- Abdo et al. 2010f, 2010g), a Narrow--Line Seyfert 1
(PKS 2004--447 -- Abdo et al., 2009b). 
The cross correlation of G10 also led to find 8 new associations among 
which two are classified as FSRQ and one as BL Lac. 
{\it Therefore most of the 1FGL--AT20G associations 
are blazars of the FSRQ or BL Lac classes.} 

The 230 1FGL--AT20G associations also have typically flat radio spectra with spectral index 
in the range $-0.5<\alpha_{(5-20\, \rm GHz)}<0.5$ and centred at $\alpha_{(5-20\, \rm GHz)}\sim 0$  
(with $F_{\nu}\propto \nu^{\alpha}$). However, the radio AT20G sources associated with a \fe\ source 
of the 1FGL catalog are only a minor fraction ($\sim$1/15) of more than 3600 AT20G sources 
with flat radio spectra (i.e. $\alpha_{(5-20\, \rm GHz)}>-0.5$). 

The 230 1FGL--AT20G associations show a statistically significant correlation between the  
$\gamma$--ray flux and the 20 GHz flux density: \fg$\propto$\fr$^{0.85\pm0.04}$ (see G10). 
This correlation has a similar slope when considering BL Lacs and FSRQs separately. 

The relevance of the \obs\ correlation is twofold: it can help to 
estimate the contribution of blazars to the 
$\gamma$--ray background (e.g. Stecker et al. 1993) and 
it could shed light on the physical link between the 
emission processes in the radio and $\gamma$--ray energy bands. 
Indeed, the so called ``blazar sequence" 
(Fossati et al. 1998, Ghisellini et al. 1998) was built by dividing 
blazars into bins of radio luminosity,
thought to be a proxy for the bolometric one, and establishes a 
link between the radio and the $\gamma$--ray emission.
On the other hand, the radio and $\gamma$--ray emitting regions are probably different,
since the rapid variability of the $\gamma$--ray flux suggests a compact size
(see e.g. Tavecchio et al. 2010),
for which the synchrotron spectrum is self--absorbed up to hundreds of GHz.
We therefore believe that the link between the radio and the $\gamma$--ray
emission (if any) must be indirect.
One possibility is that both track the jet power, with the radio averaging it
on a larger timescale than the $\gamma$--ray emission. 

The radio/$\gamma$--ray properties of EGRET blazars 
suggested the possible existence of a \obs\ correlation 
considering bright $\gamma$--ray sources (Taylor et al. 2007) 
although at low radio fluxes no conclusive claim could be made 
(e.g. Mucke et al. 1997). 
A possible correlation of \fg\ with the radio flux at 8.4 GHz  was found in the population of 
blazars detected by \fe\ in its first three months survey (LBAS sample -- Abdo et al. 2009).
This correlation was more significant for BL Lacs (chance probability $P=0.05$\%) 
than for FSRQs ($P=8$\%). 
An updated version of the \obs\ correlation, based on the \fe\ first year AGN catalog, 
is reported by Giroletti et al. (2010).
Recent studies of the radio--$\gamma$ 
flux correlation in the LBAS sources (Kovalev et al. 2009a; 2009b) was conducted using 
the MOJAVE sample of extragalactic sources (with a flux limit of 1.5 Jy at 15 GHz). 
Kovalev et al. (2009a) find that the parsec--scale radio emission and the $\gamma$--ray 
flux are strongly related in bright $\gamma$--ray objects, suggesting that \fe\ selects 
the brightest objects from a flux--density limited sample of radio--loud sources. 

The \obs\ correlation is subject to the biases related to the flux limits of the radio 
and $\gamma$--ray surveys. However, these biases acting on the radio and the 
$\gamma$--ray surveys are independent. 
The AT20G--1LAC associations have been found by cross correlating two independent surveys: the radio 
AT20G (Murphy at el. 2009) and the \fe\ 11 months survey catalogue (Abdo et al. 2010).

The two main problems we want to tackle in this paper are: 
(1) understand why only a minor fraction ($\sim$1/15) of radio sources 
(of the AT20G survey) are detected in the $\gamma$--rays by \fe, despite the 
possible existence of a correlation between the radio and the $\gamma$--ray flux; 
(2) recover the true radio--$\gamma$--ray flux correlation 
by accounting for the selection effects of both the radio and the \fe\ survey.

An important aspect which could impact on these issues is the $\gamma$--ray 
variability of blazars. We will consider in this 
paper two possible variability patterns: a long term variability which is observed, 
for instance, when comparing the fluxes measured by EGRET and 
(almost ten years later) by \fe\ for the sources in common, 
and a short term variability observed so far on daily timescales in the brightest sources.

The paper is organized as follows: in \S 2 the correlation between the $\gamma$--ray 
and the radio flux of the AT20G--\fe\ associations is presented and in \S 3 the duty 
cycle of blazars in the $\gamma$--ray band is discussed. In \S 4 we describe the 
method used to reconstruct the true $\gamma$--ray radio flux correlation and in  
\S 5 we show our results and discuss their main possible implications. 
Summary and conclusions are given in \S 6.

\section{The observed radio--gamma correlation}

\begin{figure}
\vskip -0.4 cm
\hskip -0.3 cm
\psfig{figure=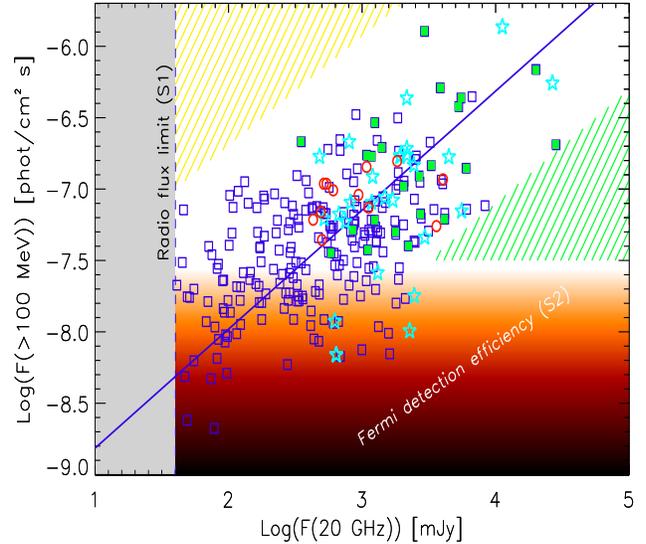,width=9.cm,height=8cm}
\vskip -0.2 cm
\caption{
The \obs\ correlation found with the 1FGL sources 
with a counterpart in the 20 GHz radio survey AT20G catalogue (open squares). 
The solid line is the fitting correlation (with slope $\sim$0.8). 
Also shown are the two main selection effects that we consider 
in this paper (see text): the radio flux limit (S1) at 40 mJy 
represented by the grey vertical shaded region and the \fe\ detection 
sensitivity which increases for increasing $\gamma$--ray fluxes 
and it is represented by the bottom shaded region. 
The two hatched triangles represent the region of the plane 
where we should expect to find sources if there is no \obs\ 
correlation. The absence of sources in these triangular regions is an 
indication of the existence of such a \obs\ correlation.  
The 3rd catalog EGRET sources (Hartmann et al. 1999)
detected also by \fe\ and not present in the 
AT20G survey (because at declination $>$0$^\circ$) are shown with open (cyan) stars,
those present in the AT20G survey by filled (green) squares and the 12 3EG sources 
not detected by \fe\ are shown with (red) open circles. 
}
\label{fg0}
\end{figure}

Fig. \ref{fg0} shows the correlation found by G10 with the 
230 1FGL \fe\ sources associated with a radio counterpart 
in the AT20G sample (squares in Fig. \ref{fg0}). 
For these associations we computed the \fe\ flux 
by integrating their spectrum (reported in the 1FGL catalogue) above 100 MeV. 
The best fit (computed with the bisector method) of the \obs\ 
correlation is \fg$\propto$\fr$^{0.8\pm0.04}$.
In the following we will refer to the latter as the 
``observed correlation" \obs.  Our aim is to account for the possible selection effects acting 
on the \obs\ plane and recover the ``real correlation" \true\ which can have a different slope and 
normalization with respect to the observed one.  
With the "hat" quantities we indicate the $\gamma$ and radio flux of the sources generated through the 
simulations described in \S 4. To these sources we apply the selection effects in order to reproduce the observed 
correlation \obs. In the simulations (see \S 4) we consider the real radio sources adopting their 
real radio fluxes (reported in the AT20G survey). Therefore, $F_{\rm r}$ and $\hat{F_{\rm r}}$ coincide, whereas for each real radio 
source, the $\hat{F_{\gamma}}$ is that obtained by assuming a certain  $\gamma$ ray variability (described in \S 3).

The two main selection effects that could bias the \obs\ correlation 
are  the flux limit of the AT20GHz survey \fr$>$40 mJy (S1) and the 
\fe\ detection efficiency in the 0.1--100 GeV energy band (S2). 
These are schematically shown in Fig. \ref{fg0}.  
While the AT20G radio survey has a well defined 20 GHz flux 
limit (shaded grey region in Fig. \ref{fg0}),
the \fe\ sensitivity depends on several parameters (Abdo et al. 2010c) 
like the source spectrum in the GeV band and its position 
in the sky where the different intensity and anisotropy of 
the galactic and extra galactic $\gamma$--ray backgrounds can 
limit the detection efficiency as a function of the source flux. 
For instance, simulations of sources distributed at high galactic 
latitudes $|b|>20^\circ$ 
show that the detection efficiency is only 1\% for sources with 
\fg\ $\sim 10^{-8}$ phot cm$^{-2}$ s$^{-1}$ although such a flux is above the 
lowest flux measured by \fe\ (Abdo et al. 2010c) . This is due to the combination of two main 
effects: the intensity of the local background and the source spectral index 
(see Abdo et al. 2010c for details). 
We show in Fig. \ref{fg0} the \fe\ detection efficiency as a shaded 
coloured region. 
This is obtained from the efficiency curve reported in Fig. 7 of Abdo et al. (2010c)
that assumes a distribution of $\gamma$--ray photon spectral indices centered
at $\Gamma=2.4$ and with a dispersion of 0.28.

Considering the distribution of sources in Fig. \ref{fg0} and the two 
instrumental selection effects S1 and S2, we note that
there are two regions in the \obs\ plane (the hatched regions in Fig. \ref{fg0}) 
where there are no sources. 
Although source number counts decrease with increasing fluxes 
(both in the radio and $\gamma$--ray band),  there seems to be no apparent instrumental 
selection effect preventing the detection of bright $\gamma$--ray sources with intermediate/low radio 
fluxes (in the yellow--hatched upper left triangle) as well as bright radio sources 
above the \fe\ detection limit (in the green--hatched lower right triangle). 
This suggests that the observed correlation is true although  we expect that its real slope and 
normalization can be different from those derived 
from the observed sources in the \obs\ plane because the latter is strongly biased 
at low \fg\ and \fr\ by the instrumental selections effects S1 and S2 (as shown in Fig. \ref{fg0}).

\section{The duty cycle of the $\gamma$--ray flux in blazars}

The variability of the $\gamma$--ray emission in blazars has been since long discussed 
in the literature. 
It is now with \fe\ that robust claims can be done, thanks to an almost continuous monitoring of 
the $\gamma$--ray sources in the sky with a relatively high sensitivity which allows to 
probe both the variations of the flux on long timescales (increasing with the mission elapsed time) 
and on short timescales (from months down to days for the brightest sources). 

Variability of the $\gamma$--ray flux of blazars could be the key ingredient to 
explain why only a minor fraction of the radio sources detected in the AT20G survey have been 
detected  by \fe\ in the $\gamma$--ray band and it could help 
to reconstruct the true \true\ correlation accounting for the \fe\ detection selection effect. 

In the following we will refer to two main variability patterns of the $\gamma$--ray flux of blazars: 
(1) a long term variability which seems to follow a log--normal probability distribution with 
$\sigma\sim$0.5 and 
(2) a short term variability which follows a non symmetric probability distribution skewed towards 
low flux levels.

\subsection{Long--timescale variability} 

In the 3rd EGRET catalog (3EG -- Hartmann et al. 1999) there are 66 high confidence AGNs. 
54 of these are detected by \fe, while 12 are not present in the 1LAC catalog. 
Among the 54 3EG sources detected by \fe, 
48 have a published radio flux density at $\sim$ 20 GHz.  Those in the northern emisphere can be added to 
Fig. \ref{fg0} (open cyan stars)  and  those in the southern emisphere (already present in the 1LAC--AT20G associations) are 
highlighted in Fig.\ref{fg0} (filled green squares).

The 48 3EG sources (out of the 54 detected by \fe), for which we could find the radio 
flux density (Fig. \ref{fg0}), are consistent with the \obs\ correlation 
found through the 1FGL--AT20G associations. 

The 12 3EG sources (classified as blazars in Hartman et al. 1999) 
not detected by \fe\ in its 11 months survey are shown in Fig. \ref{fg0} 
using their 3EG $\gamma$--ray flux (open red circles). 

The EGRET flux of these 12 sources is above the \fe\ detection sensitivity 
(shown by the shaded region in Fig. \ref{fg0}).
They occupy a region where no apparent instrumental selection effect is present. 
Therefore, the non detection of these 3EG sources by \fe\ must be due
to their $\gamma$--ray variability over about a decade. 

It is interesting to compare the $\gamma$--ray 
flux of the sources detected both by EGRET and (about 10 years after) by \fe.
When doing this, we must recall that the \fe\ fluxes are averages
over the 11 months of the survey, while the EGRET fluxes
corresponds to averages over a shorter time interval, typically
few months, since they are derived from pointed observations.

\begin{figure}
\vskip -0.4 cm
\hskip 0.3 cm
\psfig{figure=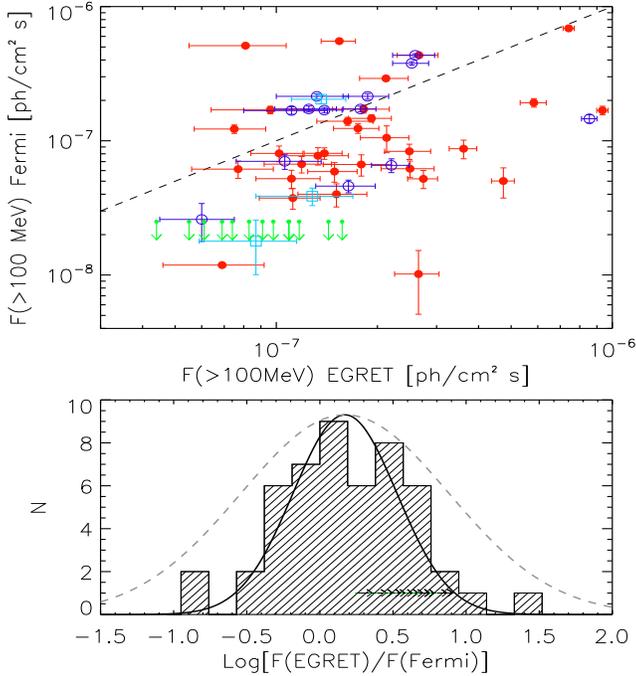,width=8.cm,height=9.5cm}
\vskip -0.2 cm
\caption{Top panel: comparison of \fe\ and 3EG fluxes ($>$100 MeV) of the  48
3EG sources detected by \fe\ (reported in the 1LAC sample)  and for which we could 
find a radio flux density at $\sim$20 GHz in the literature. The 12 3EG sources not detected 
by \fe\ are  shown as upper limits (green arrows).
FSRQs, BL Lacs and sources classified 
as ``AGN" in the 1LAC catalog are shown with different 
symbols (filled circles, open circles and open squares, respectively).  The dashed line represents equality.
Bottom panel: distribution of the ratio between the flux measured by EGRET and by \fe\ 
for the sources in common. The 12 3EG sources not detected by \fe\ are represented as lower limits 
(arrows). The  hatched distribution can be represented by a Gaussian function (solid line) with central 
value $\mu$=0.17 and standard deviation $\sigma$=0.5.  It is also shown a Gaussian (dotted grey line) with $\sigma=0.78$, 
i.e. a factor 2 larger in linear scale,  which is tested in the simulations (see \S5).
}
\label{fg00}
\end{figure}

Fig. \ref{fg00} shows the 3EG fluxes and those measured by \fe\ 
for the common sources.  We also show in Fig. \ref{fg00} the 12 3EG sources not detected by \fe\ 
as upper limits (green arrows in the top panel) and as lower limits on the 
EGRET to \fe\ flux ratio (arrows in the bottom panel of Fig. \ref{fg00}), obtained 
assuming, for \fe, a limiting flux of 2.5$\times$10$^{-8}$ ph cm$^{-2}$ s$^{-1}$, i.e. corresponding to the upper boundary of the 
shaded region S2 shown in Fig. \ref{fg0}.
On average (see the bottom panel of Fig. \ref{fg00}) the EGRET fluxes were 
larger than those of \fe. 
We find  that the ratio of the flux measured by EGRET and by \fe\ is distributed as a 
log--normal with standard deviation $\sigma$=0.5 (bottom panel of Fig. \ref{fg00}). 
We note that this distribution fully comprises also the lower limits of the 12 3EG sources not 
detected by \fe. However, given the presence of these lower limits in the bottom panel of 
Fig. \ref{fg00}, we also tested in our simulations a log--normal variability with $\sigma$=0.78 
(represented in Fig. \ref{fg00} by the dashed grey line), i.e. corresponding to 
a variation of the flux by a factor $\sim$6 .

If this is representative of a {\it decadal flux variability} of these sources, it explains
why a fraction of EGRET sources were not detected by \fe, despite its better sensitivity.

This result, i.e. the possibility that the GeV flux of $\gamma$--ray blazars 
(even if time averaged over $\sim$1 year) can vary by a factor 3 (at 1$\sigma$) over $\sim$10 years 
will be used in the next section to reconstruct the true \true\ correlation by 
accounting for the selection effect S2.

\subsection{Short--timescale variability}

Tavecchio et al. (2010) (see also Foschini et al. 2010) 
studied the  variability of the GeV flux in the two blazars 3C 454.3 (see also Bonnoli et al. 2010)
and PKS 1510--089 detected by \fe\ and found variability on few days timescale 
by considering their emission as observed by \fe\ in one year.
During exceptionally bright events, significant variability was found
also on intra--day timescales (Tavecchio et al. 2010; Abdo et al. 2010e).
Tavecchio et al. (2010) also found that, in these two sources, 
the differential flux curve (representing the number of days a 
source spends at a given flux level $F_\gamma$) has a similar pattern (also 
present in other sources -- Tavecchio et al., in preparation) 
with a rising power law and a faster decay (steep power law) bracketing a characteristic peak. 
This short--timescale variability can be described as:
\begin{equation}
N(F_{\gamma}) \propto {{(F_{\gamma}/F_{\gamma,{\rm break}})^{a} 
\cdot \exp(-F_{\gamma}/F_{\rm cut} )}
\over{1+(F_{\gamma}/F_{\gamma,{\rm break}})^{b+a} } },
\label{tav}
\end{equation}
where $F_{\gamma,{\rm break}}$ is the flux corresponding to the break 
between the low--flux power law with slope $a$ and the high--flux 
power law with slope $b$ and $F_{\rm cut}$ is the flux of the exponential cutoff. 
Since we are concerned with $\gamma$--ray fluxes {\it averaged} over one year,
the short--timescale variability can produce {\it very modest} variations of
the averaged flux.

Instead, the long--timescale variability observed in the common EGRET/\fe\ sources can 
change the flux by a factor 3 (at 1$\sigma$ level), implying 
a larger spread of the $\gamma$--ray flux, although corresponding to 
longer timescales.

\section{The simulation}

We want to constrain the normalization and slope of the real \true\ correlation 
that reproduces the observed distribution of the 
230 1LAC--AT20G associations shown by squares in Fig. \ref{fg0}
and, at the same time, accounts for the non--detection of the large majority of the radio sources
of the AT20G sample.
Even if the $\gamma$--ray detection rate approaches 100\% at the largest radio fluxes and
decreases for lower radio fluxes,
this is not a trivial task.
This is because there are many radio sources, undetected by \fe, with a radio flux
comparable or even larger than those that are instead detected in $\gamma$--rays.

By accounting for the \fe\ detection efficiency and for the assumed duty cycle of blazars in the 
$\gamma$--ray band, we search for the \true\ correlation that produces a distribution of 
simulated sources in the \obs\ plane which matches the observed one.  

In particular, we consider all the radio sources with flat radio spectrum in 
the AT20G survey and assign to them a $\gamma$--ray flux according to a given 
\true\ correlation (whose normalisation 
and slope are the free parameters that we want to constrain). 
Then we shuffle the $\gamma$--ray flux of each source according to a law which is 
representative of the $\gamma$--ray variability and apply the \fe\ detection sensitivity 
to identify those simulated sources that should be detected by \fe. 
In this way we populate the \obs\ plane with simulated sources observable by \fe\ for any 
assumed \true\ correlation. 
We constrain the slope and normalisation of the \true\ correlation by requiring that  
(i) the number of sources that should be detected by \fe\ is consistent with the 
real number of sources defining the \obs\ correlation (i.e. 230$\pm$15), 
(ii) that the distribution of \fr\ and \fg\ of 
the simulated sources is consistent with those of the real sources (this is evaluated through 
the Kolmogorov--Smirnov test, performed separately, between the \fr\ and \fg\ distributions of the 
simulated and of the real sources). 
The details of the simulation and its assumptions are described below,
and we present our results in \S~5.

\begin{figure*}
\hskip 1.5  cm
\psfig{figure=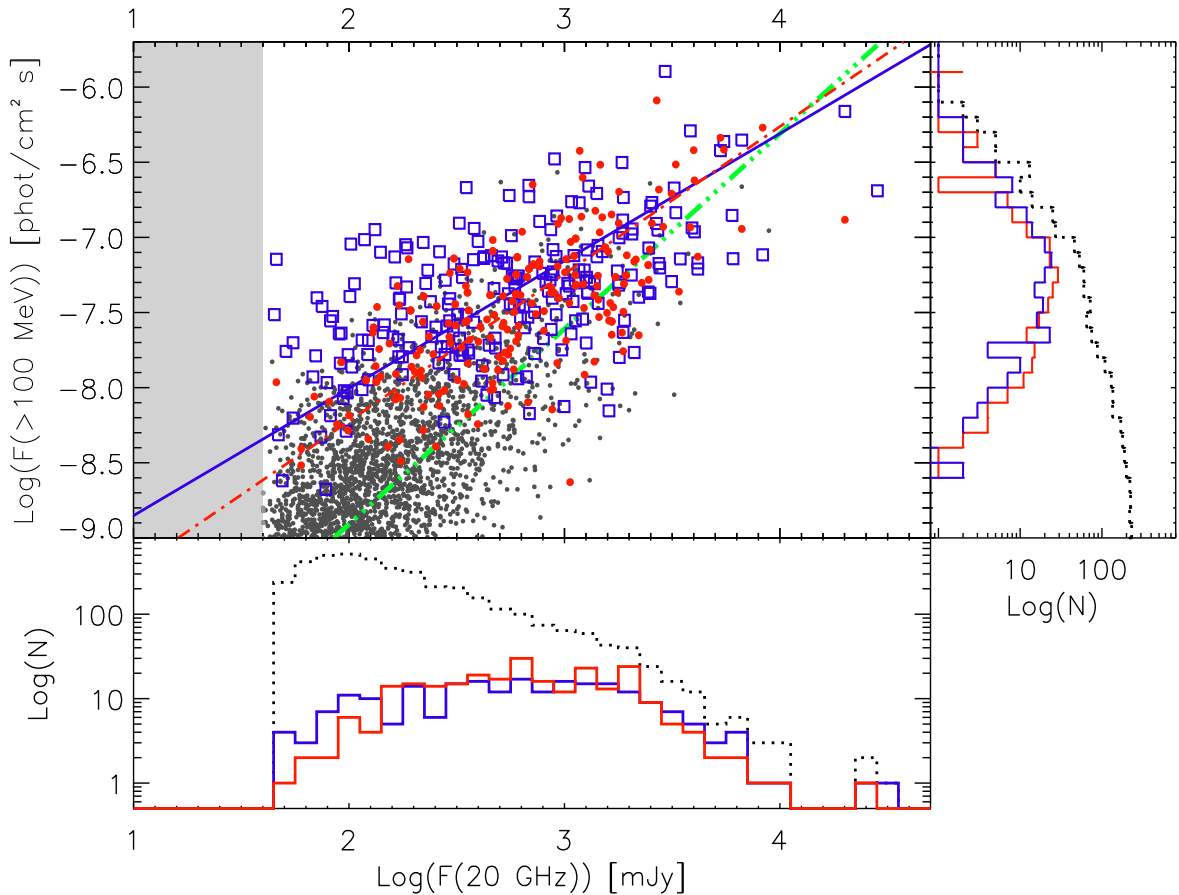,width=16.3cm,height=12cm}
\caption{
Example of a simulation of sources following the true 
correlation \true\ (triple dot--dashed green line) with  $\hat{\delta}=1.3$ and 
obtained with a $\gamma$--ray  log--normal variability distribution
with $\sigma_{\gamma}=0.5$. 
Filled black dots are the real AT20G sources with flat 
radio spectra ($\alpha_{(5-20\, \rm GHz)}>-0.5)$ simulated with the log--normal 
\fg\ variability function. 
The red points are 250 simulated sources which can be detected by \fe\ according to its 
detection efficiency curve (selection effect S2 -- Abdo et al. 2010c). 
The dot--dashed line is the best fit correlation of simulated sources  
 and the solid blue line is the best fit to the real sources. The open blue squares are the 230 real sources 
of G10. The bottom and right--hand side panels show the \fr\ and \fg\ 
distributions of the total simulated sample (dotted black line)  and 
of the simulated sources after the application of the selection 
effects (red line). The distribution of the fluxes of the real sources 
detected by {\it Fermi} is shown by the blue line. 
}
\label{fg1}
\end{figure*}

The simulation relies on some input assumptions: 
\begin{enumerate}
\item 
the true \true\ correlation is modelled as a power law 
\begin{equation}
\log(\hat{F_{\gamma}})=\hat{K}+\hat{\delta} \log(\hat{F_{r}}) 
\label{fgfr}
\end{equation}
where the normalisation $\hat{K}$ and the slope $\hat{\delta}$ 
are the free parameters that we aim to constrain. The normalization 
is computed at 150 mJy in our simulations. 
This particular value corresponds to the average of the radio fluxes of the 
AT20G sources with flat radio spectrum used for the simulations;

\item we consider the 3686 radio sources of the AT20G survey with flat radio spectrum, 
i.e. $\alpha_{(5-20\, \rm GHz)}>-0.5$, similar to the radio spectrum of the 230 1FGL--AT20G associations 
defining the observed \obs\ correlation;

\item the $\gamma$--ray flux variability: 
we assign to each radio source 
with a certain $\hat{F_{\gamma}}$ (given by Eq. \ref{fgfr}) a flux 
\fg\ according to one of 
the two possible variability functions described in \S 3. 
First we assume the short--timescale variability function, described by  Eq.\ref{tav}.
Since we are concerned with averaged (over 11 months) fluxes, we extract 11 $\gamma$--ray
fluxes from the $N(F)$ distribution of Eq. \ref{fgfr} after having fixed its parameters to
$F_{\gamma,{\rm break}}=\hat{F_{\gamma}}$ and always setting
$a=1.5$, $b=3$ and $F_{\rm cut}=5\, F_{\gamma,{\rm break}}$. 
We then average the 11 values of \fg\ obtained in this way for each simulated source.
The obtained flux is different from the initial $\hat{F_{\gamma}}$, but by a small factor,
and we anticipate that the 
dispersion induced by this treatment of variability is much smaller than the dispersion of the real 
sources in the \obs\ plane along the \fg\ axis. 

For this reason we adopted, as a second choice, the long--term variability function, i.e. a log--normal 
distribution with assigned standard deviation.
This assumption is motivated by the comparison of the EGRET and \fe\ flux for the common
sources shown in Fig. \ref{fg00} (see \S 3). 
The bottom panel of Fig. \ref{fg00} shows that the distribution of the flux ratio 
has a standard deviation of 0.5 which we also assume in our simulation. However, we will 
test a log--normal variability function with larger/lower standard deviation.

We stress that while the variability function of Eq. \ref{tav} is representative of the 
blazar activity over one year,  the log--normal with $\sigma=0.5$ 
found in \S 3 corresponds to a variability over ten years at least, i.e. 
the time between the EGRET and the \fe\ measurement of the {\it average} flux of the sources in common.  
Furthermore, it is already indicative of how the $\sim$1 year {\it average} $\gamma$--ray flux
varies, not of the variations occurring on shorter timescales, as instead indicated by Eq. \ref{tav}.
For this reason, we will extract only one flux from this log--normal variability distribution.

\end{enumerate}

We did not model the possible radio variability of blazars. This is motivated by the fact that we simulate the $\gamma$--ray flux of 
real radio sources, i.e. those with flat radio spectra in the AT20G complete survey. These are 3686 sources with \fr$\ge$ 40 mJy:
their large number ensures that we are sampling the possible range of variability of the radio flux density. 

\begin{figure}
\vskip -0.4 cm
\hskip -0.4 cm
\psfig{figure=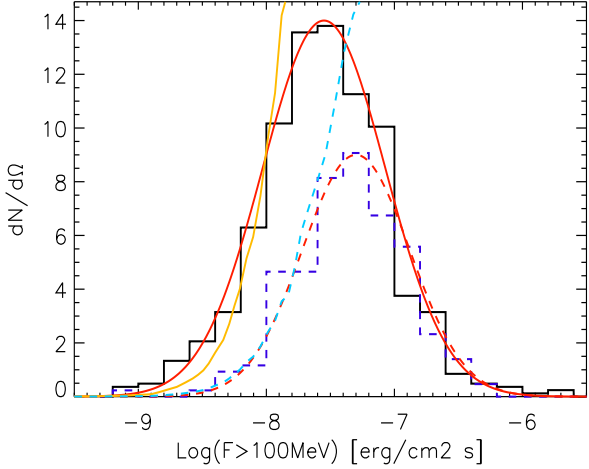,width=9.cm,height=7cm}
\vskip -0.2 cm
\caption{Flux distribution of \fe\ sources of the 1LAC sample located at high galactic latitudes 
($|b|>20^{\circ}$ -- solid histogram) and located 
along the Galactic plane ($|b|<20^{\circ}$ -- dashed blue histogram). 
The two distributions are normalized to the corresponding solid angles. 
The solid and dotted red lines are two Gaussians fitting the flux distributions. 
The dashed (cyan) and solid (orange) lines show the detection efficiency (adapted from Abdo et al. 2010c) 
and scaled by a factor 3 between the two histograms. 
}
\label{fg000}
\end{figure}

\subsection{One illustrative example}

In Fig. \ref{fg1} we show an example of a simulation. 
For this example we have assumed a \true\ correlation with 
slope $\hat{\delta}=1.3$ and $\hat{K}=-8.6$ at 150 mJy.
After having assigned to each radio source its corresponding $\hat{F_{\gamma}}$,
we have extracted its $F_\gamma$   
from a log--normal distribution peaking at $\hat{F_\gamma}$ and having a dispersion 
$\sigma=0.5$.
The sources simulated in this way are shown by the grey dots in Fig. \ref{fg1} and 
the assumed \true\ correlation  is shown by the triple--dot--dashed (green) line. 
Then we selected the sources detectable by \fe\ (red dots) with the following procedure.
We have considered the  
\fe\ $\gamma$--ray detection efficiency (selection effect S2) which 
is a function of the $\gamma$--ray flux (shaded area in Fig. \ref{fg0}). 
For S2 we used the detection efficiency curve presented in Fig. 7 
of Abdo et al. (2010c) which was obtained through simulations of sources 
at galactic latitudes $|b|>20^\circ$. 

For sources at low galactic latitudes the level of the galactic 
background and the larger number of sources may reduce the detection efficiency. 
Therefore, we mimic this effect by considering a detection 
efficiency reduced by a factor of 3 for sources at $-20<b<20^{\circ}$. 
This choice is motivated by Fig. \ref{fg000}
where it is shown the $\gamma$--ray flux distribution of the \fe\ sources of the 
1LAC sample located at high galactic latitudes ($|b|>20^{\circ}$ -- solid histogram) 
and located along the Galactic plane ($|b|<20^{\circ}$ -- dotted blue histogram). 
The two curves matching the left--hand side of the histograms represent the detection 
efficiency rescaled by a factor 3 for the sources at low galactic latitudes. 

Among the simulated sources in a given bin of $\gamma$--ray flux we randomly 
extract a fraction of sources corresponding to the 
\fe\ detection efficiency (from Abdo et al. 2010c) in that flux bin. 
This corresponds to the application of the S2 selection bias. 
The simulated sources surviving the S2 selection (i.e. ``detectable sources", hereafter) 
are the red dots in Fig. \ref{fg1}. 

Their distributions in the \obs\ plane is compared 
with the distribution of real sources in the same plane. 
First, we compare independently the distributions of \fg\ and \fr\ of the 
detectable and real sources through the Kolmogorov--Smirnov test and 
derive the corresponding probabilities $P(KS)_{\gamma}$ and $P(KS)_{r}$. 
The two histograms (of the detectable and real sources) are shown in the 
two side--panels of Fig. \ref{fg1}.
We consider that the detectable and the real sources have similar distributions 
in the \obs\ plane when the KS probabilities are both $>10^{-2}$. 

Then, for each assumed \true\ correlation (with fixed slope and normalization, 
$\hat{\delta}$ and $\hat{K}$)
we repeat the simulation 300 times and count the number of simulations yielding a 
KS probability $>$1\% that {\it both} the radio and $\gamma$--ray flux distributions of 
detectable and real sources are drawn from the same parent population. 

We consider a set of input parameters ($\hat{\delta}$, $\hat{K}$) acceptable when more 
than 68\% of the 300 simulations had $P(KS)_{\gamma}$ and $P(KS)_{r}$  larger than 1\%. 

 Finally, among the acceptable simulations we identified those 
producing a number of detectable sources  (red points in Fig. \ref{fg1})  equal to the real one (i.e. 230$\pm$15, open 
blue squares in the example of Fig. \ref{fg1}). 
In the example shown in of Fig. \ref{fg1} 
the number of simulated sources detectable by \fe\ is $\sim$250. 
These simulations give us the slope and normalization of the true
\true\ correlation we are seeking.

\section{Results}

In the following section we present the results obtained through our simulations under the 
two possible variability scenarios for the $\gamma$--ray flux discussed in \S. 3.

\subsection{Simulations with the short--term $\gamma$--ray variability}
In Fig. \ref{fg2} we show the number of simulated sources 
detectable by \fe\ versus the slope $\hat{\delta}$ of the \true\ correlation.
These are the results obtained assuming a short--term variability of the 
$\gamma$--ray flux described in \S. 3.2.
Each curve represents a different normalization $\hat{K}$ of the \true\ correlation. 
The open grey squares in Fig. \ref{fg2} are those simulations rejected because 
the distribution of \fg\ and/or \fr\ of the simulated sources 
are inconsistent with those of the real sources, i.e. $P(KS)_{\gamma}$ 
and/or $P(KS)_{r}$ $<$ 10$^{-2}$, in more than 68\% of the repeated simulations. 

Vice versa, the filled circles 
correspond to  distributions of simulated sources 
(i.e. the red points in Fig. \ref{fg1}) in the \obs\ plane consistent with 
the distribution of the real sources, i.e. in more than 68\% of the repeated 
simulations (for each choice of the free parameters $\hat{\delta}$ and $\hat{K}$)  
the $P(KS)_{\gamma}$ and $P(KS)_{r}$ $>$ 10$^{-2}$.

The results shown in Fig. \ref{fg2} are obtained under the hypothesis 
that the short time variability of the $\gamma$--ray flux is described by Eq. \ref{tav},
implying a very modest variation of the average flux.
We note that all the acceptable simulations (filled circles) over predict the number 
of sources with respect to the 230 \fe\ detected sources with an AT20G counterpart (G10). 
Therefore a modest variability of the average $\gamma$--ray flux
cannot reproduce the number of sources really observed in the \obs\ plane for any 
assumed true \true\ correlation.

\subsection{Simulations with the long--term $\gamma$--ray variability}

\begin{figure}
\hskip -0.4 cm
\psfig{figure=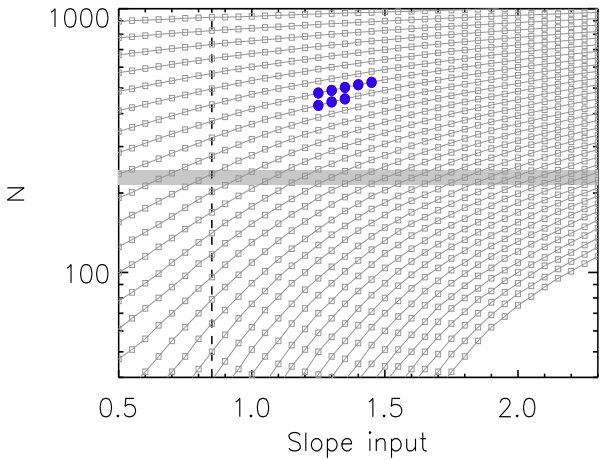,width=9.cm,height=7cm}
\caption{
Simulation results. 
The number of simulated sources (which survive the two instrumental 
selection effects S1 and S2 described in the text) is plotted against 
the slope $\hat{\delta}$ of the assumed \true\ correlation. 
Each curve represents a set of simulations of the \true\ correlation 
with fixed normalization $\hat{K}$ and varying slope $\hat{\delta}$. 
The open squares are those cases of ($\hat{\delta}$, $\hat{K}$) where 
more than 68\% of the simulations fail to reproduce the observed source 
distribution of the real sources in the \obs\ plane (i.e.  in more than 68\% 
of the simulations the simulated--source \fg\ and \fr\ distributions 
have KS probabilities $<$1\% of being consistent with the \fg\ and \fg\ 
distributions of the real sources). The acceptable simulations are shown 
by the filled blue circles. 
The simulations are performed assuming Eq. \ref{tav} for the $\gamma$--ray variability 
function
that, after averaging, results in a very modest flux variability.
The shaded region represents the number of real sources (i.e. 230$\pm$15) 
detected by \fe\ with a radio counterpart which give rise to the 
observed \obs\ correlation. For reference, the vertical dashed line shows
the slope of the observed \obs\ correlation. 
 }
\label{fg2}
\end{figure}

\begin{figure}
\hskip -0.4 cm
\psfig{figure=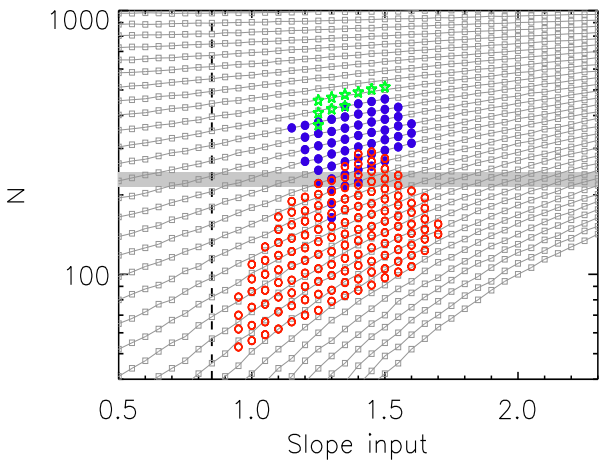,width=9.cm,height=7cm}
\caption{
Same as Fig. \ref{fg2} but with a variability function for the $\gamma$--ray flux 
which is a log--normal with $\sigma_{\gamma}$=0.5 (filled blue circles). 
Also shown are the results assuming a 
$\gamma$--ray log--normal variability function with $\sigma_{\gamma}$=0.18 
and $\sigma_{\gamma}$=0.78 (shown 
by the open green stars and the open red circles respectively).
}
\label{fg3}
\end{figure}

The next step was to assume  a larger variability function for \fg, i.e. a 
log--normal distribution with $\sigma_{\gamma}=0.5$. 
The results of the simulations under this assumption for the variability of 
\fg\ are shown in Fig. \ref{fg3} (filled blue circles). 
In this case the acceptable \true\ correlations extend over a wider 
parameter range of normalization and slope, $\hat{K}$ and 
$\hat{\delta}$, of the \true\ correlation with respect to the 
acceptable simulations shown of Fig. \ref{fg2}. 
This is due to 
the assumed larger amplitude variability.
In this case there is a set of simulations (those intersecting the horizontal 
shaded stripe in Fig. \ref{fg3}) that can also reproduce the real number 
of observed sources. 
Therefore the solutions we find in this case correspond to  
$-8.7<\hat{K}<-8.3$ and $1.25<\hat{\delta}<1.5$.


The results shown in Fig. \ref{fg2} on the slope and normalisation 
can be understood with the aid of Fig. \ref{fg1}. 
The distribution of real sources in the \obs\ 
plane (open blue squares in Fig. \ref{fg1}) is constraining. 
For instance, for very low normalisations $\hat{K}$ 
of the \true\ correlation, 
the number of simulated points surviving the 
selection effect S2, i.e. the red points in Fig. \ref{fg1}, 
is lower than 230 and their distributions 
of \fg\ and \fr\ are inconsistent with that of the real sources. 
This is the case of the simulations below the shaded stripe (representing the 
number of real sources in the \obs\ plane) in Fig. \ref{fg2}. 

However, when a log--normal 
function is assumed for the variability of \fg, the spread of the simulated points in \fg\ can 
be larger and, although the 
detectable sources are much fewer
than the real one, 
their \fg\ and/or \fr\ distributions can still be consistent with those of the real sources. 
This explains why there are solutions in Fig. \ref{fg3} which are acceptable although the number 
of detectable sources is smaller than the real one. 

Although the choice of a log--normal variability function with $\sigma_{\gamma}=0.5$ 
is motivated by the long term variability of the EGRET sources detected by \fe, 
we also verified how the solutions of the simulation depend on the choice of $\sigma_{\gamma}$. 
In particular we tested a log--normal variability function with $\sigma_{\gamma}=0.18$ 
and $\sigma_{\gamma}=0.78$ corresponding to a linear flux variation by a factor 
1.5 and 6 respectively. 
The solutions are shown in Fig. \ref{fg3} by the open star and circles, respectively. 
We find that $\sigma_{\gamma}=0.18$ introduces a too small degree of variability 
(similar to the variability function of Eq. \ref{tav}) 
and all the solutions over predict the 
number of detectable sources with respect to the real number of associations. 
On the other hand a larger variability (i.e. $\sigma_{\gamma}=0.78$) extends the 
space of acceptable solutions below those obtained with $\sigma_{\gamma}=0.5$. 
 Note that the position of the red open circles and of the green open stars in Fig. \ref{fg3} (corresponding to 
simulations with $\sigma_{\gamma}=0.78$  and $\sigma_{\gamma}=0.15$, respectively) does not exactly coincide with 
that of the filled blue circles. This is because the three sets of simulations, shown in Fig. \ref{fg3}, have slightly different normalizations 
even when the slope is equal. For clarity, in Fig. \ref{fg3} we draw only the curves (open grey connected squares) corresponding to the 
simulation with $\sigma_{\gamma}=0.5$.

As a caveat we stress that in our simulations, we adopted the \fe\ sensitivity computed by 
Abdo et al. (2010c) which assume a spectral index distribution typical of FSRQ sources. 
While it is known that the \fe\ sensitivity strongly depends on the source 
spectral index (e.g. Abdo et al. 2010a), it should be noted that our sample of 1FGL--AT20G associations is dominated by 
FSRQ sources. For BL Lac objects the better sensitivity of \fe\ in detecting these hard sources would imply a lower detection limit 
(represented by the shaded region S2 in Fig.\ref{fg0}). This would require, in order to reproduce the observed \obs\ correlation, a   combination of a slightly smaller normalization and slope of the \true\ correlation possibly coupled with a slightly larger variability of the $\gamma$--ray flux. Still the results would be comparable with those derived with the detection sensitivity of FSRQ since we find that a variability of at least a factor 3 (i.e. $\sigma$=0.5) is necessary to reproduce the \obs\ correlation.

\subsection{Predicted number of {\it FERMI} detectable sources}

The \fe\ sensitivity is increasing with the increasing 
exposure time of its survey. 
We can use the simulation and the reconstructed \true\ correlation to infer 
the number of sources that will be detected  with a future increase 
of the survey time which will improve the 11 months \fe\ survey limit by a factor of 2. 
We expect that a larger number of detected $\gamma$--ray sources will have a 
counterpart in the AT20G radio survey. 
By running our simulation with the best correlation found in the 
previous section, we find that 
the total number of southern sources present in ATG20G that will be detected
by {\it Fermi} will go from the current 230 to $\sim$430. 
This number is in agreement with what expected if the radio $LogN-LogS_{\rm r}$ has slope --3/2 and 
considering the reconstructed correlation $\hat{F_{\gamma}}\propto \hat{F_{\rm r}}^{1.5}$,
and implies a $\gamma$--ray $LogN-LogS_{\gamma}$ with a slope flatter than euclidean.

\subsection{Gamma--Radio luminosity correlation}

The possible correlation between the radio and the $\gamma$--ray luminosity 
has been studied in the past with the aid of EGRET detected sources. 
Different groups reported a significant correlation between the radio luminosity 
(at frequencies larger than 1 GHz) and the $\gamma$--ray one 
(e.g. Fossati et al. 1998, Salamon \& Stecker 1996) of blazars detected by EGRET. 
Bloom (2008) found $L_{\gamma}\propto L_{r}^{0.77\pm0.03}$ with a sample of 122 sources 
identified as blazars in the revised EGRET sample. 
Mucke et al. (1997) argued that the correlation could be due to instrumental 
biases coupled to the use of average $\gamma$--ray fluxes that washes out the considerable 
variability of blazars at $\gamma$--ray  wavelengths. 
Nontheless, Zhang et al. (2001), through partial correlation analysis, showed 
that a marginal correlation exists between the radio and the $\gamma$--ray 
luminosity in EGRET blazars considering the high and low state fluxes of EGRET 
sources nearly simultaneously observed in the radio band. 
More recently, Pushkarev et al. (2010) showed that there 
exists a strong correlation between the $\gamma$--ray and the 
VLBA radio flux on monthly timescales and that the radio flux lags 
the $\gamma$--ray one by 1--8 months.

\begin{table}
\centering
\begin{tabular}{lllll}
\hline
\hline
Sources             &$L_\gamma$--$L_r$ &$L_\gamma$--$z$ &$L_r$--$z$   &$L_\gamma$--$L_r$ $(z)$ \\
\hline
All (144)           &0.8                &0.89            &0.72         &0.5        \\
                    &10$^{-34}$         &10$^{-45}$      &10$^{-25}$   &10$^{-11}$  \\
FSRQs (112)         &0.66	               &0.83            &0.54         &0.44       \\
                    &10$^{-15}$         &10$^{-30}$      &10$^{-10}$   &10$^{-7}$     \\
BL Lacs (22)        &0.78	               &0.90            &0.78         &0.29       \\
                    &10$^{-5}$          &10$^{-9}$       &10$^{-5}$    &10$^{-1}$     \\
AGNs (10)           &0.96               &0.97            &0.94         &0.56          \\
                    &10$^{-6}$          &10$^{-6}$       &10$^{-5}$	    &10$^{-1}$ \\		
FSRQs+BL Lacs (134) &0.76	               &0.86            &0.67         &0.5       \\
                    &10$^{-27}$         &10$^{-42}$      &10$^{-18}$   &10$^{-10}$     \\
\hline
\hline
\end{tabular}
\vskip 0.4 true cm
\caption{Partial correlation analysis of the $L_{\gamma}$--$L_{r}$ correlation 
accounting for the common redshift dependence. 
Each line gives the Spearman correlation coefficients and in the 
last column it is reported the partial correlation coefficient. 
The probabilities of the correlation coefficient are also given. 
}
\label{tab1}
\end{table}

\begin{figure}
\hskip -0.9 cm
\psfig{figure=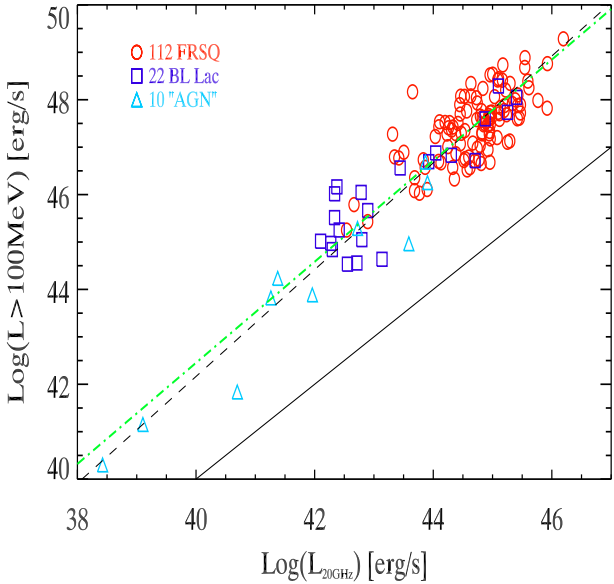,width=9.8cm,height=9cm}
\caption{K--corrected $\gamma$--ray luminosity (integrated above 100 MeV) versus radio 
luminosity ($\nu L_\nu$ at 20 GHz) for 
the 112 FSRQs, 22 BL Lacs and 10 ``AGNs" (different symbols as shown in the legend) of the AT20G--1LAC 
association sample with measured redshifts. The solid line represents equality. 
The dashed line is the fit considering all the sources (slope 1.13) and the dot--dashed 
line is the fit considering only FSRQs and BL Lacs (slope 1.07).
 }
\label{fg5}
\end{figure}

Among the 230 AT20G--1LAC associations there are 144 sources 
(112 FSRQs, 22 BL Lacs and 10 ``AGNs") with measured redshifts. 
In Fig. \ref{fg5} we show them in the $\gamma$--ray versus radio luminosity plane. 
Both luminosities have been k--corrected using the radio and 
$\gamma$--ray spectral index of individual sources.  
While the $\gamma$--ray luminosity is integrated above 100 MeV, the radio one is 
the $\nu L_\nu$ luminosity computed at  20 GHz. 
FSRQs, BL Lacs and ``AGNs" are distributed along a linear correlation. 
Considering only FRSQs and BL Lacs (circles and squares in Fig. \ref{fg5}) the correlation has a 
slope 1.07$\pm$0.05 (dot--dashed line in Fig. \ref{fg5}). 
A somewhat steeper slope 1.13$\pm$0.04 is found if also ``AGNs" are included in the 
fit (i.e. considering all 144 sources)\footnote{All 
the fits are performed with the bisector method (e.g. Isobe et al. 1990).}.
This value is steeper than that found by Bloom (2008) and also of the correlations 
(both for the low/high states and for the 
average flux case of EGRET blazars) reported by Zhang et al. (2001). 

Given the common dependence of the $\gamma$--ray and radio luminosity 
on the redshift $z$ we should test if the \lglr\ correlation is true. 
Several methods have been applied to investigate this possibility (e.g. Mucke et al. 1997, 
Zhang et al. 2001, Bloom et al. 2008). 
We perform a partial correlation analysis by removing the dependence of both 
$L_{\gamma}$ and $L_{\rm r}$ on the redshift $z$ (e.g. Padovani 1992).  
We computed the Spearman correlation coefficients and the associated probabilities 
and then the partial correlation coefficient and the probability of the null hypothesis 
that the two luminosities are uncorrelated. 
The chance probability of the partial correlation coefficient is distributed as a t--statistic. 
All the values of the correlation coefficients and the associated probabilities are reported in Tab. 1. 
We note that considering the FSRQs 
and BL Lacs together the partial correlation probability of the null hypothesis is 10$^{-10}$. 
This result indicates, in agreement with that reported by Bloom (2008),  
that indeed a strong \lglr\ exists in blazars. 
However, by considering BL Lacs and AGNs separately, we find a high chance probability 
of the partial correlation coefficient. 
This suggests that, although we still have few sources 
of these classes, their large redshift spread makes the luminosity correlation less statistically 
significant than for the class of FSRQs.

\subsection{Contribution of blazars to the EGBR}

The existence of a \lglr\ correlation and of a corresponding correlation in the observer frame 
(i.e. the \true\ correlation found in this paper) could also have some implications for the computation 
of the contribution of blazars to the extragalactic $\gamma$--ray background radiation (EGBR). 
One method often adopted to this aim uses a linear relation between the $\gamma$--ray luminosity 
of blazars and the luminosity at some other wavelength in order to re--scale the $\gamma$--ray 
luminosity function through the often better known luminosity function at the other wavelength 
(e.g. Salamon \& Stecker 1996, Norumoto \& Totani 2006). Alternatively one can construct the 
$\gamma$--ray luminosity function of blazars starting from a catalog, like the \fe\ first 
blazar catalog. 
The latter method has been recently applied by Abdo et al. (2010c). 

\fe\ finds (Abdo et al. 2010d) that the EGBR spectrum is consistent with a power law with spectral 
index 2.41$\pm$0.05 and an integrated ($>$100 MeV) flux of 1.03$\times 10^{-5}$ cm$^{-2}$ 
s$^{-1}$ sr$^{-1}$, which is softer and less intense with respect to the measurements of EGRET.
Note that the above value is not the total background, but the one obtained 
subtracting out the detected sources.
It contains the contribution of undetected sources which have either a flux below the \fe\ sensitivity 
threshold (corresponding to the flux of the faintest source detected by \fe) or that are 
not detected (but with a flux larger than this limit) because of their intrinsic properties 
(e.g. soft spectrum or position in the sky coincident with regions of high diffuse background 
level -- see Abdo et al. 2010c). 
Recently, Abdo et al. (2010c) considered the contribution of blazars to the EGBR. 
They point out that, due to the detection efficiency of \fe\ (that we also used in this work), 
there is a substantial fraction of $\gamma$--ray sources which are not detected but still have 
a flux larger than the flux of the faintest source detected by \fe. 
Therefore, from their count distribution, they estimate that non--detected blazars, but with 
a flux larger than the faintest detected source limit, should contribute the $\sim$16\% 
of the EGBR flux (to which detected \fe\ sources have been subtracted). 
By extrapolating the blazars' count distribution to zero flux, this estimate becomes 
23\% (Abdo et al. 2010c). 

We can perform a simple exercise: we assume that all the radio sources in the AT20G 
survey with flat radio spectrum are candidate blazars emitting in the \fe\ energy band. 
Through the \true\ correlation we can compute their integrated \fg\ that can be compared 
with the level of the $\gamma$--ray background. 
This estimate should be compared with the EGBR flux including the detected sources, 
i.e. roughly a factor 1.3 larger than the EGBR flux used in Abdo et al. (2010c) 
from which the detected sources were removed. The diffuse EGBR we use is adapted 
from Fig. 3 of Abdo et al. (2010d) where the EGBR and the contribution of detected 
sources are shown separately. 

We use the reconstructed \true\ correlation with slope in the range 1.25--1.4 
(corresponding to the acceptable solutions in Fig. \ref{fg3} -- filled circles) and 
perform a set of 300 simulations first assigning to the 3686 flat radio spectrum AT20G 
sources an $\hat{F_\gamma}$ through this assumed correlation and then assigning
an $F_\gamma$ according to a 
log--normal $\gamma$--ray flux variability with $\sigma_{\gamma}=0.5$ (\S 4). 
For each simulation we calculate the total $\gamma$--ray flux contributed by 
the flat radio sources. 
On average we find that their contribution (according to 
the assumed correlation slope and normalization) ranges between the $\sim$37$\pm5$\%  
and the $\sim$52$\pm5$\%  (for slopes 1.25 and 1.4, respectively) of the 
EGBR including the detected sources. 
This fraction is made by two contributions: 
the total flux of detected sources (i.e. the 230 AT20G--\fe\ associations) which can 
be between 20\% and 35\% of the EGBR, while the remaining 17\% is the contribution 
of undetected sources belonging to the population of flat radio sources with \fr$\ge$40 mJy 
(i.e. in the AT20G survey). 
Based on model population studies of blazars, Inoue et al. (2010) 
finds that the contribution of blazars to the EGBR (including detected sources) should be 45\% 
(an additional 35\% should be due to non--blazar AGNs). 
This  estimate is consistent with the range derived from our analysis. 
We note that in our estimate we are considering the combined 
contribution of FRSQs and BL Lacs, although they have different $\gamma$--ray spectral properties 
(the latter, having a harder spectrum in the \fe\ band, are very likely dominating the 
contribution to the EGBR at high energies) and redshift distributions. 
However, we have 
used the \true\ correlation for our estimate  and we have shown (G10) that both these 
classes of sources follow a similar correlation between the radio and the $\gamma$--ray flux.

\subsection{Predictions for the brightest $\gamma$--ray blazars}

\begin{figure}
\hskip -0.6 cm
\psfig{figure=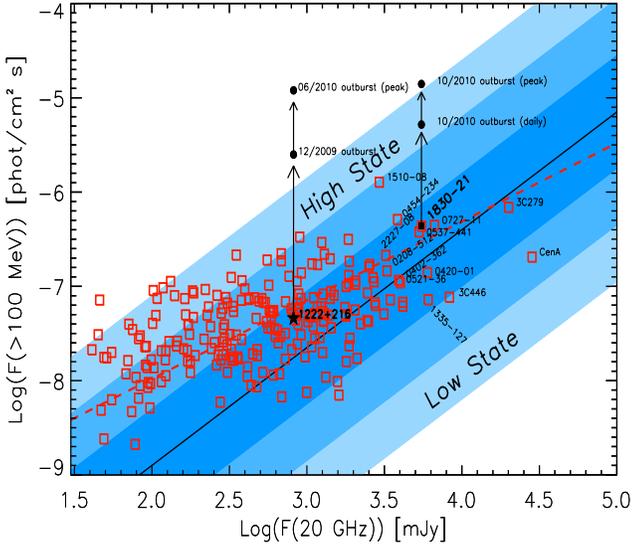,width=9.2cm,height=8.cm}
\caption{$\gamma$-ray flux versus radio flux density of the AT20G--1LAC 
associations (open red squares). 
The solid line represents the reconstructed \true\ correlation
and its 1, 2 and 3 $\sigma_{\gamma}$ scatter (with $\sigma_{\gamma}=0.6$) is shown by 
the shaded regions. The names of the brightest sources are shown. 
The region above the solid line (with slope 1.25) represents the high state: 
i.e. a source can vary its average \fg\ within the shaded regions according 
to a log--normal probability function with assigned $\sigma$. 
The dashed line is the \obs\ correlation. 
The outbursts detected by \fe\ of two sources, i.e. PKS 1222+216 and PKS 1830-21, are shown.
}
\label{fg6}
\end{figure}

One possible application of  the \true\ correlation obtained in this paper is to 
predict the average $\gamma$--ray flux and its maximum value 
for any given blazar with known radio flux density at 20 GHz. 
Fig. \ref{fg6} shows the AT20G--1LAC associations (G10) 
and the reconstructed correlation (here we have chosen to report 
the solution with slope 1.25) and its 1, 2 and 3 $\sigma_{\gamma}$ dispersion. 
The \true\ correlation divides the plane of Fig. \ref{fg5} into two regions that 
we label as ``low" and ``high" state. 
These correspond to those states of the $\gamma$--ray flux (1, 2 or 3 $\sigma$) above or 
below the average value represented by the \true\ correlation (solid line in Fig. \ref{fg6}). 
The shaded regions in Fig. \ref{fg6} show how the average flux (over 11 months as measured 
by \fe\ in its survey which is the one we used to construct the 
\obs\ correlation) can vary according to a log--normal distribution with $\sigma_{\gamma}\sim$0.6.

We have considered the brightest blazars, with \fr\ at $\sim$ 20 GHz larger than 3 Jy, 
distributed in the southern (i.e. present in the AT20G survey) and northern emisphere. 
For them we can calculate the average $\hat{F_\gamma}$ and the maximum average \fg\ they can reach 
if their long--term variability follows the log--normal distribution 
found for the sources in common between EGRET and \fe. 
These values of $\hat{F_\gamma}$  and \fg$_{,3 \sigma}$ are given in Tab. \ref{tab2}. 
Among the sources in which the $\gamma$--ray 
flux can be larger than 5$\times 10^{-5}$ ph cm$^{-2}$ s$^{-1}$
there are 3C 279 and 3C 273. 
Furthermore, Tavecchio et al. (2010) showed that in the brightest blazars a short--term variability 
(\S 3) can be present and modulate the $\gamma$--ray flux by a factor 3 (or even 10 
in the most extreme cases), on top of the long term variability. 
Therefore, they could reach even larger fluxes through sporadic 
flares of emission, like the case of 3C 454.3 last year (Bonnoli et al. 2010; Foschini et al. 2010), 
increasing their \fg\ from the highest average value (i.e. the \fg$_{,3 \sigma}$) still by a 
factor of a few. 
Tab. \ref{tab2} provides the list of those sources which can be included 
in long--term multi--wavelength monitoring of blazars for the study of their variability 
and for the characterization of the most extreme phases of their emission.

An interesting case is represented by 4C+21.35 (PKS 1222+216 at z=0.43). 
In April 2009 this source increased its average \fg\ flux by a factor of 
10 with respect to its average flux of $\sim$4.6$\times 10^{-8}$ phot cm$^{-2}$ s$^{-1}$ 
measured in the \fe\ first 6 months survey (Longo et al., 2009). 
A strong flare was detected by AGILE/GRID in December 2009 (Verrecchia et al. 2009) 
with a flux (integrated above 100 MeV) of 2.5$\times 10^{-6}$ phot cm$^{-2}$ s$^{-1}$ 
and confirmed by \fe\ (with flux 3.4$\times 10^{-6}$ phot cm$^{-2}$ s$^{-1}$ -- Ciprini et al., 2009). 
In the period April--May 2010 GeV flares were detected by \fe\ (Donato et al., 2010) 
at a flux level of 8$\times 10^{-6}$ phot cm$^{-2}$ s$^{-1}$, i.e. a factor about 4 
in excess with respect to the average flux of that period and by AGILE (Bulgarelli et al., 2010). 
Very High Emission (above 100 GeV) was also found from this source in this period 
(Neronov et al., 2010; Mariotti et al., 2010). 
Finally in June 2010 \fe\ recorded a 
flux of 1.2$\times 10^{-5}$ phot cm$^{-2}$ s$^{-1}$ (Iafrate et al., 2010) which 
represents an increase of a factor of 3 with respect to the average flux of the week.  

4C+21.35 has a radio flux (at 15 GHz) of about 1 Jy. 
The \true\ correlation can be used to 
calculate its average flux level and its average 3$\sigma$ flux level. 
These turn out to be $\hat{F_{\gamma}}$=2.2$\times 10^{-8}$ phot cm$^{-2}$ s$^{-1}$ and 
\fg$_{, 3\sigma}$=1.4$\times 10^{-6}$ phot cm$^{-2}$ s$^{-1}$. 
We note that the brightest flare of 4C+21.35 detected by \fe\ in June 2010, 
has a flux a factor $\sim$10 larger than the maximum average flux predicted 
from the \true\ correlation, i.e. if the source were 3$\sigma$ brighter than 
its average flux as predicted by the correlation. 

As already discussed, one possibility is that while the average flux (i.e. the average as measured by \fe\ in 
its 11 months survey) is modulated by a long--term variability following a log--normal 
distribution as found in the case of the EGRET sources detected by \fe\ (see \S 3), 
on top of this there is a much shorter variability (described by Eq. \ref{tav}) 
as found in the brightest blazars (Tavecchio et al. 2010). 
This short--timescale variability operates on top of each state of the 
average $\gamma$--ray flux and can boost the flux still by a factor 3--10 
even when the source has reached its maximum value of the average flux.

In Fig. \ref{fg6} we show the variations of the $\gamma$--ray flux of PKS 1222+216 caught to 
be in outburst by both \fe\ and AGILE on December 2009 (Verrecchia et al. 2009, Ciprini et al. 2009)
and to reach in June 2010 the flux of 1.2$\times 10^{-5}$ phot cm$^{-2}$ s$^{-1}$. We also report the 
recent detected outburst of PKS 1830-21 (Ciprini et al. 2010) which went through an outburst in October 2010 reaching
a daily \fg\ of 5.2$\times 10^{-6}$ phot cm$^{-2}$ s$^{-1}$ with a peak flux of 1.4$\times 10^{-5}$ phot cm$^{-2}$ s$^{-1}$.

\begin{table}
\centering
\begin{tabular}{lccc}
\hline
\hline
Source        & \fr &  $\hat{F_{\gamma}}$   &  \fg$_{,3\sigma}$   \\
		 & Jy   & phot cm$^{-2} s^{-1}$ &  phot cm$^{-2} s^{-1}$ \\
\hline
3C 279	       		  & 20.0     &	9.5e-7	&	6.0e-5  \\
$[$HB89$]$ 1921-293	  & 13.8  	&	6.0e-7	&	3.8e-5	\\	
3C 446	       		  & 8.3      	& 3.2e-7		& 2.0e-5	\\
PKS 0727-11	       	  & 6.7  	&	2.4e-7	&	1.5e-5 	\\
PKS 1335-127           & 6.1      	&2.1e-7		&1.4e-5	\\
PKS 0420-01	          & 6.0      	&2.1e-7		&1.3e-5	\\
PKS 1830-21	          & 5.5      	&1.9e-7		&1.2e-5	\\
PKS 0537-441           &  5.3      &	1.8e-7	&	1.1e-5	\\
PKS 0454-46	         &  4.2      	&1.3e-7		&8.5e-6	\\
CRATES J0609-0615    & 4.2       &	1.3e-7	&	8.5e-6	\\
PKS 0402-362         &  4.0    	&	1.2e-7	&	8.0e-6	\\
PKS B0607-157        &  4.0      	&1.2e-7		&8.0e-6	\\
PKS 0521-36	        &   4.0      	&1.2e-7		&8.0e-6	\\
PKS 0454-234         &   3.8  	&	1.2e-7	&	7.5e-6	\\
PKS 1954-388         &   3.8      	&1.2e-7		&7.5e-6	\\
AP LIB	             &    3.4  	&	1.0e-7	&	6.5e-6	\\
PKS 0208-512         &   3.3      	&1.0e-7		&6.3e-6	\\
PKS 2227-08	        &    3.2      	&9.6e-8		&6.0e-6	\\
PKS 0637-75	        &    3.1      	&9.2e-8		&5.8e-6	\\
PKS 1510-08	        &    2.9      	&8.5e-8		&5.3e-6	\\
\hline
 3C 273 		        &   23.8	&	1.2e-6	&	7.4e-5 \\
 3C 345 		        &   12.0	&	5.0e-7	&	3.2e-5 \\
 3C 454.3		       &   11.0	&	4.5e-7	&	2.3e-5 \\
87GB[BWE91] 0059+5808  & 8.6	&	3.3e-7	&	2.1e-5 \\
 $[$HB89$]$ 2145+067	 &  8.5	&	3.3e-7	&	2.1e-5 \\
 OJ +287		         &  6.0	&	2.1e-7	&	1.3e-5 \\
$[$HB89$]$ 0735+178	 &  5.3	&	1.8e-7	&	1.1e-5 \\
$[$HB89$]$ 2134+004	 &  5.1	&	1.7e-7	&	1.1e-5 \\
$[$HB89$]$ 0923+392 	 &  5.0	&	1.7e-7	&	1.1e-5 \\
 BL Lac 		         &  4.5	&	1.5e-7	&	9.2e-6 \\
$[$HB89$]$ 2201+315	 &  4.5	&	1.5e-7	&	9.2e-6 \\
4C +50.11		 &  4.2	&	1.3e-7	&	8.4e-6 \\
$[$HB89$]$ 1055+018 	 &  4.2	&	1.3e-7	&	8.5e-6 \\	
$[$HB89$]$ 1308+326	 &  3.9	&	1.2e-7	&	7.6e-6 \\
$[$HB89$]$ 1611+343	 &  3.6	&	1.1e-7	&	6.9e-6 \\
$[$HB89$]$ 1928+738	 &  3.5	&	1.1e-7	&	6.7e-6 \\
LBQS 0106+0119 	 &  3.5	&	1.1e-7	&	6.8e-6 \\
$[$HB89$]$ 2005+403	 &  3.4	&	1.0e-7	&	6.4e-6 \\
$[$HB89$]$ 0234+285	 &  3.4	&	1.0e-7	&	6.5e-6 \\
$[$HB89$]$ 0642+449	 &  3.3	&	1.0e-7	&	6.4e-6 \\
$[$HB89$]$ 1749+096	 &  3.3	&	1.0e-7	&	6.3e-6 \\
\hline
\hline
\end{tabular}
\vskip 0.4 true cm
\caption{Blazars with the highest radio flux density at 20 GHz, i.e. \fr$\ge$3 Jy, 
distributed in the southern and northern emisphere (top and bottom table, respectively). 
For each source it is reported its name, the radio flux density (for southern sources this is extracted from the AT20G survey - Murphy et al. 2009) while for the northern 
sources it is taken from the Nasa Extragalactic Database and in most cases it is at 22 GHz. We give the average $\gamma$--ray flux \fg calculated from the reconstructed \true\ correlation (i.e. solid line in Fig.~\ref{fg6}) and its 3 $\sigma$ highest value corresponding to the highest possible state according to a log-normal variability of the $\gamma$--ray flux with $\sigma_{\gamma}$=0.6 (shaded region in Fig.~\ref{fg6}).  
}
\label{tab2}
\end{table}

\section{Discussion and conclusions}

We studied the \obs\ correlation between the $\gamma$--ray ($>100$ MeV) flux and the 
radio flux (at 20 GHz) observed in the  230 \fe\ 
sources with a counterpart in the 20 GHz ATCA survey (G10). 
This correlation is biased by the radio flux limit of the AT20G survey 
and at low $\gamma$--ray fluxes by the \fe\ sensitivity. 
However, in the \obs\ plane there are regions (hatched triangles in 
Fig. \ref{fg0}) where 
these selection effects are not present and still no source is found. 
This suggests that the \obs\ correlation is real. 
And yet, only 1/15 of the radio sources with flat radio 
spectra in the AT20G survey have a counterpart in the 11 months \fe\ survey.

Through numerical simulations we have recovered the true \true\ 
correlation that can reproduce the observed \obs\ one. 
In doing this we have considered the two main instrumental selection effects 
(radio flux limit and \fe\ sensitivity) and 
we tested the possibility
that the non detection of radio sources by \fe\ could be 
due to the variability of their emission in the $\gamma$--ray energy range.
 Tavecchio et al. (2010) characterized the variability of 3C 454.3 
and PKS 1510--089, two among the brightest blazars detected by \fe\ (see also Foschini et al. 2010). 
They found a daily flux distribution which we here modeled through Eq. \ref{tav}. 
However, this ``short" timescale variability induces a small variation of the 
$\gamma$--ray flux 
when we average it over 1 year.
A longer timescale variability is found by comparing,
for the sources in common, the fluxes measured by EGRET and, after almost ten years, by \fe.
The ratio of the $\gamma$--ray fluxes of the blazars detected by EGRET 
and \fe\ have a log--normal distribution with a standard deviation $\sigma_{\gamma}=0.5$, 
and 12 of the 66 blazars detected by EGRET have not been detected by \fe. 
This suggests that the ``short" timescale variability is superimposed 
on a ``decadal" variability and that the latter has a larger relative variation. 
We considered these two variability patterns in our simulations aimed at recovering the 
true \true\ correlation of the AT20G--\fe\ associations. 

As shown in Fig. \ref{fg2}, we cannot reproduce the observed \obs\ correlation if we adopt 
the short timescale variability 
(and average the flux over 1 year).
In this case we over predict the number of simulated sources detectable 
by \fe\ for any combination of the \true\ correlation slope and normalization. 
Instead, we find (Fig. \ref{fg3}) that the distribution of real sources 
in the \obs\ plane can be reproduced if the slope of the true \true\ correlation is 
in the range 1.3-1.5 and if the $\gamma$--ray 
(1--year averaged)
flux variability is modeled as a 
log--normal with $\sigma_{\gamma}\ge$ 0.5 (acceptable solutions in Fig. \ref{fg3} represented by 
filled and open circles). We also find that a smaller flux variability, also  modeled as 
a log--normal, cannot reproduce the observed \obs\ correlation, similarly to what found
by assuming the short timescale variability pattern of Eq. \ref{tav}.
 
Therefore, we need a ``decadal" $\gamma$--ray variability which modulates the 
1--year averaged $\gamma$--ray flux by at least a factor 3 in order to explain the radio/$\gamma$--ray 
statistical behavior of blazars. 
The need to assume such a variability pattern is corroborated by the findings
on the  ratio of EGRET to \fe\ flux for the sources in common, although this is based on the 
EGRET sources which are the most luminous (and maybe the most variable). However, our results show that 
at least a variability of a factor 3 in flux is necessary to reproduce the observed radio--$\gamma$ flux correlation. 
Recent results, obtained after the 11 months \fe\ survey, are 
illuminating: PKS 1222+216 increased its average $\gamma$--ray flux by a factor
300 with respect to its average flux during the first 6 months \fe\ survey. 
This implies that not only the $\gamma$--ray flux can vary on very short timescales, 
but also on longer timescales and with an amplitude that is larger than what seen in the radio. 
 
For our simulations, we we used the \fe\ sensitivity computed by Abdo et al. (2010c) which assumes a spectral index distribution 
typical of FSRQ sources. Although  the \fe\ sensitivity depends on the source spectral index (e.g. Abdo et al. 2010a),  our sample of 1FGL--AT20G associations is dominated by 
FSRQ sources. Nonetheless, for BL Lac objects, due to their harder spectrum, a lower \fe\ detection limit 
(represented by the shaded region S2 in Fig.\ref{fg0}) should be considered for these sources. We remark that based only on the radio data we cannot distinguish in our simulations between FSRQ and BL Lac objects among the radio sources with flat radio spectrum but without a $\gamma$--ray counterpart. However, if a lower detection sensitivity would be used in our simulations, a   combination of a slightly smaller normalization and slope of the \true\ correlation possibly coupled with a slightly larger variability of the $\gamma$--ray flux would be required to reproduce the \obs\ correlation. Still the results would be comparable with those derived with the detection sensitivity of FSRQ.

As a result of the existence of the \true\ correlation and of the long--timescale 
variability we also expect that the brightest radio blazars (e.g. 
with radio flux density $\ge$3 Jy) will always be detected by \fe\ since the 
variability of the $\gamma$--ray flux cannot make them much fainter
(on average, short duration events at very low fluxes can occur).
Our result also implies that if the average fluxes in the radio and 
$\gamma$--ray bands are correlated, then also a decadal 
radio flux variation should be expected and with a positive time delay  
if the radio emitting
region is larger than the one emitting the $\gamma$--rays, and thus is located 
further out along the jet.

However, our fluxes are collected at different epochs:
the radio data are single snapshots obtained by the AT20G survey in the period 2004--2008
while the \fe\ fluxes are averages over 11 months survey between 2008 and 2009.
Pushkarev et al. (2010) have shown that there exists a strong correlation between the 
15.4 GHz core VLBA flux density and the $\gamma$--ray one and that these two fluxes 
are correlated with a typical delay 
distributed between 1 and 8 months (in the observer frame) which is interpreted as 
caused by opacity effects. 

This study, on the correlation of {\it fluxes}, is related to the possibility 
that there exists a correlation between the 
$\gamma$--ray and the radio {\it luminosities} in blazars (as found e.g. by Bloom et 
al. 2008 and discussed in Mucke et al. 1997). 
Although based on the fraction (144) of the AT20G--\fe\ associations with 
measured redshifts, we showed that a strong correlation 
exists between the radio luminosity and the $\gamma$--ray one and 
that, considering FSRQs and BL Lac objects, it is linear. We also verified, 
through partial correlation analysis, that this correlation is not due to 
the common dependence of the luminosities on redshift for FSRQs 
(null hypothesis probability of the partial correlation, removing 
the redshift dependence, $P=10^{-10}$), while only a marginal claim 
can be  made for BL Lacs alone.

The other main consequence of the existence of a \true\ correlation in 
blazars is that it could be used to estimate the 
contribution of these sources to the $\gamma$--ray background (Abdo et al. 2010c). 
The contribution of blazars can be estimated considering the 
true \true\ correlation and the population of radio sources with flat 
radio spectra (i.e. candidate blazars) in the AT20G survey. 
We estimate that the blazar contribution to the extragalactic diffuse background 
is between 37\% and 52\% (according to the parameters of the assumed \true\ correlation) 
of which 17\% is the contribution from non--detected sources.  Considering that our 
estimates are based on a flux limited sample of radio sources (i.e. those with 
20 GHz flux larger than 40 mJy) we should expect that if the radio flux limit 
is further decreased, the number of sources should increase. 
While detailed predictions depends on the still poorly known $LogN-LogS$ 
of radio sources at very low flux levels, we should expect that 
the contribution of blazars to the EGRB can even be larger than our 
present estimates based on a radio flux limited sample.

\section*{Acknowledgments}
 We thank the referee for useful comments and suggestions. 
This work was partly financially supported by a 2007 COFIN-MIUR grant and 
ASI I/088/06/0 grant.


\begin{thebibliography}{}
\bibitem[]{} Abdo, A. A., et al.,  2009, ApJ 700, 597
\bibitem[]{} Abdo A.A. et al., 2009a, ApJ, 707, L142
\bibitem[]{} Abdo, A. A., et al., 2010, ApJS, 188, 405 (A10)
\bibitem[]{} Abdo, A. A., et al., 2010a, ApJ 715, 429 (A10a)
\bibitem[]{} Abdo A. A., et al., 2010b, ApJ, 709, L152
\bibitem[]{} Abdo A. A., et al., 2010c, ApJ subm., arXiv:1003.0895 
\bibitem[]{} Abdo A. A., et al., 2010d, Phys. Rev.  Subm., arXiv:1002.3603 
\bibitem[]{} Abdo A. A., et al., 2010e, ApJ subm., arXiv:1007.0483
\bibitem[] {} Abdo A. A.,  et al., ApJ, 2010f, 719
\bibitem[] {} Abdo A. A., et al., ApJ, 2010g, accepted, arXiv:1006.5463
\bibitem[]{} Atwood, W. B., et al., 2009, ApJ, 697, 1071
\bibitem[]{} Bonnoli, G. et al., 2010, arXiv:1003.3476
\bibitem[]{} Bloom, S. D., 2008, ApJ, 136, 1533
\bibitem[]{} Bulgarelli et al., 2010, ATel, \#2641
\bibitem[]{} Ciprini S., et al., 2009, ATel \#2349
\bibitem[]{} Ciprini S., et al., 2010, ATel \#2943
\bibitem[]{} Donato, D., et al., 2010, ATel \#2584
\bibitem[]{} Foschini et al., 2010,  MNRAS, 408, 448
\bibitem[]{} Fossati G., Maraschi L., Celotti A., Comastri A. \& Ghisellini G.,
              1998, MNRAS, 299, 433
\bibitem[]{} Ghirlanda G., et al., 2010, MNRAS in press, arXiv1003.5163 (G10)
\bibitem[]{} Ghisellini G., Celotti A., Fossati G., Maraschi L. \&
              Comastri A., 1998, MNRAS, 301, 451
\bibitem[]{} Ghisellini, G., Tavecchio F., \& Maraschi, L., 2009, MNRAS, 396, L105
\bibitem[]{} Ghisellini, G. \& Tavecchio, F., 2009, MNRAS, 397, 985
\bibitem[]{} Giroletti, M. et al., 2010, 009 Fermi Symposium, arXiv:1001.5123
\bibitem[]{} Gliozzi M. et al., 2008, A\&A, 478, 723
\bibitem[]{} Hartmann et al. 1999, ApJS, 123, 79
\bibitem[]{} Iafrate, G. et al., 2010, ATel \#2687
\bibitem[]{} Inoue, Y., et al., 2009, Proc. of the Fermi Symp., arXiv:1001.0103
\bibitem[]{} Isobe T., et al., 1990, ApJ, 364, 104
\bibitem[]{} Longo, F. et al., 2009, ATel, \#2021
\bibitem[]{} Kovalev, Y. Y., et al. 2009a, ApJ, 707, L56
\bibitem[]{} Kovalev, Y. Y., et al., 2009b, ApJ, 696, L17
\bibitem[]{} Mahony, E., et al., 2010, ApJ Subm., arXiv:1003.4580
\bibitem[]{} Mariotti, M., et al., 2010, ATel \#2684
\bibitem[]{} Murphy, T, et al., 2010, MNRAS, 402, 2403
\bibitem[]{} Mucke et al. 1997, A\&A, 320, 33
\bibitem[]{} Neronov D., et al., 2010, ATel \#2617
\bibitem[]{} Norumoto, T. \& Totani, T., 2007, Astrophys. Space Sci, 309, 73 
\bibitem[]{} Padovani, P., 1992, A\&A, 256, 399
\bibitem[]{} Pushkarev, A. B., Kovalev, Y. Y., Lister, M. L., 2010, ApJ Subm., arXiv:1006.1867
\bibitem[]{} Salamon, M. H. \& Stecker, F. W., 1996, ApJ, 430, 21
\bibitem[]{} Stecker F. W., Salamon  M. H., Malkan  M. A., 1993, ApJ, 410, L71
\bibitem[]{} Tavecchio F. et al., 2010, MNRAS, 401, 1570
\bibitem[]{} Taylor et al. 2007, ApJ, 671, 1355
\bibitem[]{} Zhang, L., Cheng, K. S. \& Fan, H., 2001, PASJ, 53, 207
\bibitem[]{} Verrecchia, F., et al., 2009, ATel \#2348
\end{thebibliography}
\end{document}